\documentclass[
,final            
]
{aipproc}
\usepackage{amssymb}
\usepackage{graphicx}
\usepackage{subfigure}

\layoutstyle{8x11single}

\makeatletter
\def\ScaleIfNeeded{%
  \ifdim\Gin@nat@width>\linewidth
  \linewidth
  \else
  \Gin@nat@width
  \fi
}
\makeatother

\begin{document}

\title{Numerical Simulations of Driven Supersonic Relativistic MHD Turbulence}


\author{Jonathan Zrake and Andrew MacFadyen}{address={Center for Cosmology and Particle Physics,
    Physics Department, New York University, New York, NY 10003} }

\begin{abstract}
  Models for GRB outflows invoke turbulence in relativistically hot
  magnetized fluids.  In order to investigate these conditions we have
  performed high-resolution three-dimensional numerical simulations of
  relativistic magneto-hydrodynamical (RMHD) turbulence.  We find that
  magnetic energy is amplified to several percent of the total energy
  density by turbulent twisting and folding of magnetic field lines.
  Values of $\epsilon_B \gtrsim 0.01$ are thus naturally expected.  We
  study the dependence of saturated magnetic field energy fraction as
  a function of Mach number and relativistic temperature.  We then
  present power spectra of the turbulent kinetic and magnetic
  energies.  We also present solenoidal (curl-like) and dilatational
  (divergence-like) power spectra of kinetic energy.  We propose that
  relativistic effects introduce novel couplings between these
  spectral components.  The case we explore in most detail is for
  equal amounts of thermal and rest mass energy, corresponding to
  conditions after collisions of shells with relative Lorentz factors
  of several.  These conditions are relevant in models for internal
  shocks, for the late afterglow phase, for cocoon material along the
  edge of a relativistic jet as it propagates through a star, as well
  neutron stars merging with each other and with black hole
  companions.

  We find that relativistic turbulence decays extremely quickly, on a
  sound crossing time of an eddy.  Models invoking sustained
  relativistic turbulence to explain variability in GRB prompt
  emission are thus strongly disfavored unless a persistant driving of
  the turbulence is maintained for the duration of the prompt
  emission.
\end{abstract}

\maketitle


\section{Introduction}
Turbulent cascades in astrophysics are invoked to explain a range of phenomena,
including the modification of galactic magnetic fields \citep{Kulsrud:1992p3916}
and star formation rates \citep{Shu:1987p3935, MacLow:1999p3923,
  Mckee:2007p3924}. In this regime of gasdynamical turbulence it is safe to use
the equations of isothermal magnetohydrodynamics, because the relevant flows are
slow, cold, and weakly magnetized. However, magnetic fields may undergo
turbulent amplification in environments where these assumptions are grossly
violated. For example, the amplification of magnetic fields at neutron star
mergers is driven by the Kelvin-Helmholtz instability \citep{Price:2006p3270,
  Obergaulinger:2010p2801} operating in an extremely hot and strongly magnetized
turbulent medium. Also, gamma-ray burst (GRB) outflows involve
ultra-relativistic bulk flows containing mildly relativistic (warm) thermal
velocities, \citep{Medvedev:1999p483, Goodman:2008p3936} and require weak
magnetic fields to account for observed synchrotron radiation. Due to the
additional coupling between inertial dynamics and gas pressure introduced by
relativistic effects, existing paradigms for compressible MHD turbulence cannot
be employed to understand these environments.
\begin{figure}
  \centering
  \includegraphics[width=1.7in]{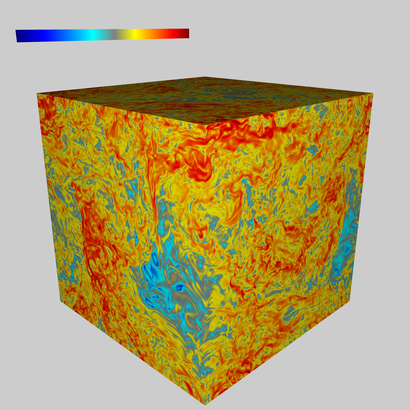}
  \includegraphics[width=1.7in]{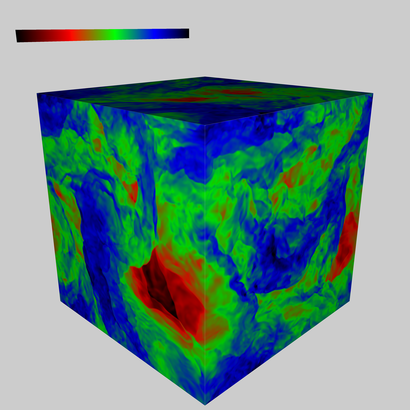}
  \includegraphics[width=1.7in]{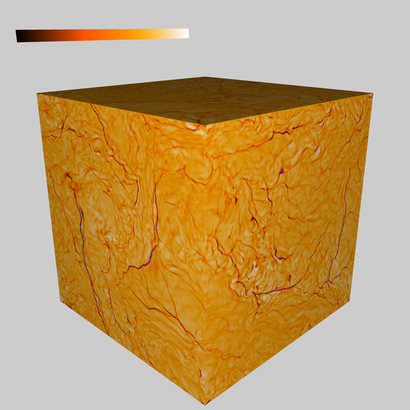}
  \includegraphics[width=1.7in]{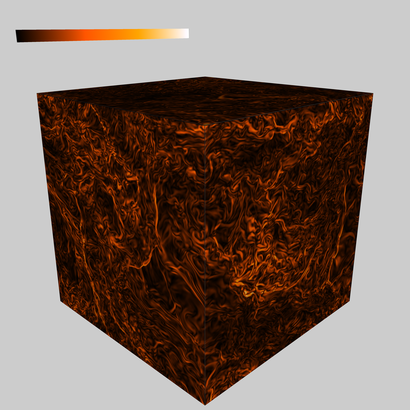}
  \caption{Shown are surface renderings of the magnetic energy
    (\emph{top left}), gas pressure (\emph{top right}), the vorticity
    (\emph{bottom left}) and the divergence (\emph{bottom right}) of
    the velocity field. This snapshot was taken after 2 light-crossing
    times, which is roughly the time at which $\epsilon_B$ becomes
    saturated. The volume averaged Mach number at this time is $\sim
    1.7$. This model was run at resolution $512^3$ using the forcing
    parameter $F_0=6.0$.}
  \label{fig:3dbox-all-sm}
\end{figure}

The mechanism for prompt emission of gamma-ray bursts (GRBs) is a major open
problem in astrophysics. The internal shock model \citep{Piran:1993p4065,
  Katz:1994p4165, Rees:1994p4101} has been widely applied to modeling GRB prompt
emission. Recently however, \cite{Narayan:2009p4019} have suggested an
alternative model in which variability in the GRB prompt emission is produced by
relativistic fluctuations in the velocity field in the frame of the
outflow. \cite{Kumar:2009p4009} claim that observations of GRB080319B rule out an
internal shock model and suggest that the prompt emission is produced near the
deceleration radius ($\sim 10^{17} \rm{cm}$) by emitters with random Lorentz
factors of ($\Gamma \sim 10$) in the comoving frame of the
outflow. \cite{Lazar:2009p4221} have computed light curves from simplified models
and conclude that the observed light curves can be obtained in a turbulence or
``mini-jet'' model if the emitter size and the bulk and random Lorentz factors
satisfy tight constraints which may be inconsistent with other light curve
features. In these models the light curve variability comes from the random
Lorentz factor. On physical grounds, however, fluid turbulence should have
random fluctuations of the velocity field on scales of the sound speed $(\Gamma
\sim 1)$ since supersonic flow will rapidly shock and dissipate. Detailed
simulations of the dissipation of supersonic turbulence is therefor of interest
in evaluating the plausibility of novel models for GRB emission.

\section{Description of Turbulence models}
Our simulations are carried out on the 3-dimensional periodic cube
$[-0.5,0.5]^3$. All velocities are measured in units of $c$, and thus time is
measured in units of the light-crossing time of the box. We initialize a uniform
fluid of density $\rho_0=1.0$ having a background seed field $\mathbf{B}_0 = B_0
\hat{\mathbf{x}}$ such that $\epsilon_B = 10^{-5}.$ The relativistic
temperature, which we will define by $T \equiv P/\rho c^2$ is kept fixed
throughout this study at $T = 1/3$. This value is deliberately chosen so that
the fluid's rest mass energy density and internal energy density are in rough
equipartition.
\begin{figure}
  \centering
  \includegraphics[width=3.0in]{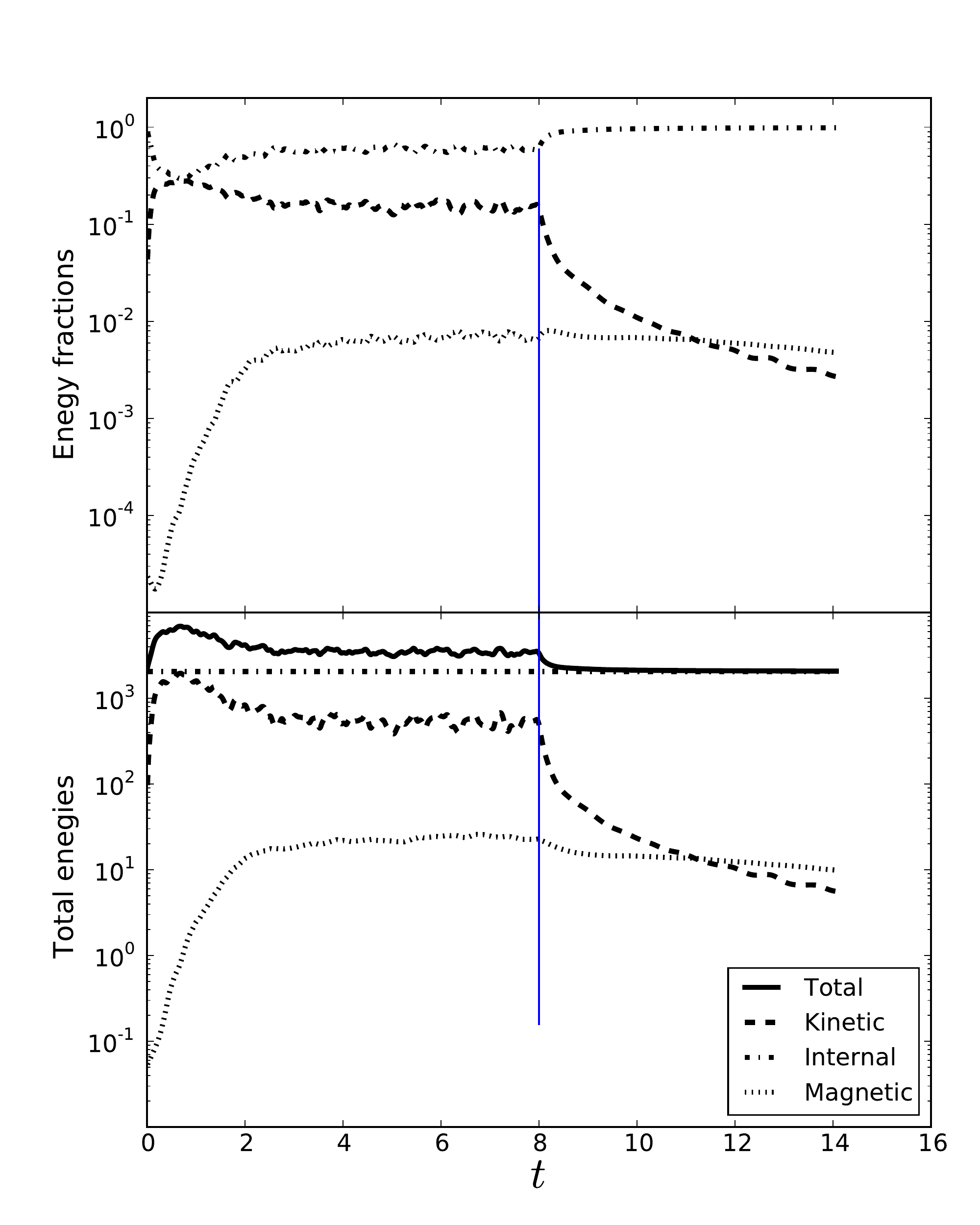}
  \includegraphics[width=3.0in]{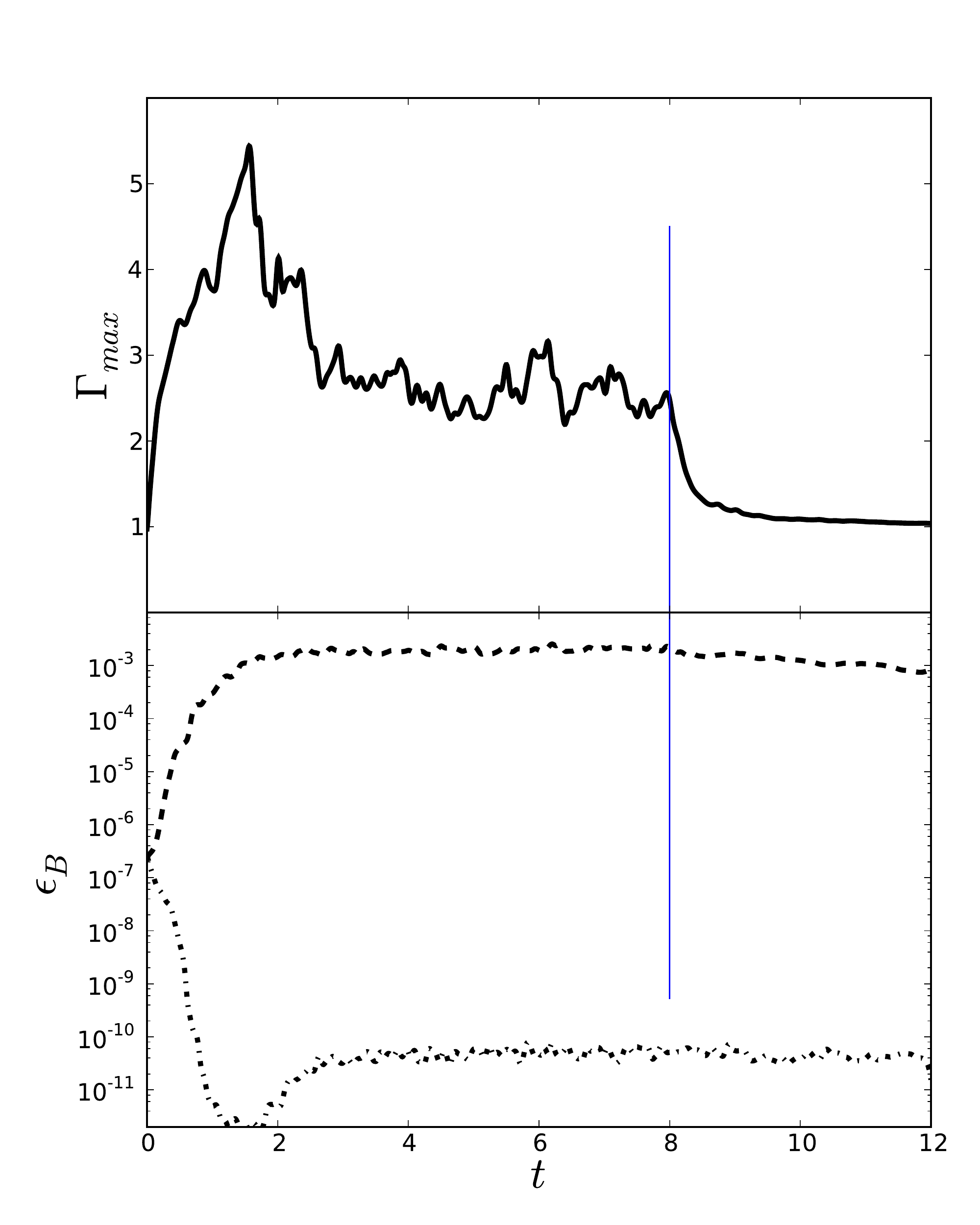}
  \caption{\textbf{Left}: Time histories of the total (\emph{solid}), kinetic
    (\emph{dashed}), inernal (\emph{dash-dotted}), and magnetic (\emph{dotted})
    energies for supersonic turbulence. \textbf{Right}: Maximum Lorentz factor
    and the maximum (\emph{dashed}) and average (\emph{dash-dotted}) magnetic
    energy fraction. In these models, driving is switched off at $t=8$ marked by
    the vertical line.}
  \label{fig:1111-512-EnergyGrowth}
\end{figure}

We use the same driving procedure in all runs, with the exception of the RMS
forcing parameter $F_0$ which is adjusted to obtain the desired Mach number. We
have adopted the forcing parameters $F_0=0.25$ and $F_0=6.0$ for subsonic and
supersonic models respectively. Driving is switched on at the start of the
simulation and kept on for $8$ light-crossing times, after which driving is
switched off and the flow is left to decay. For our supersonic turbulence model,
we explore two branches of decay in the absence of driving, both starting from
the same conditions at the moment driving is switched off. In the first branch,
the cooling procedure is kept in place, maintaining the effectively isothermal
euquation of state. In the second branch, we explore the free decay of adiabatic
turbulence. Note that the adiabatic equation of state becomes practical again
for decaying turbulence since the injection of energy has ceased, and thus the
increase of pressure due to viscous heating at the grid scale is limited by the
conservation of total energy.
\begin{figure}
  \centering
  \includegraphics[width=3.0in]{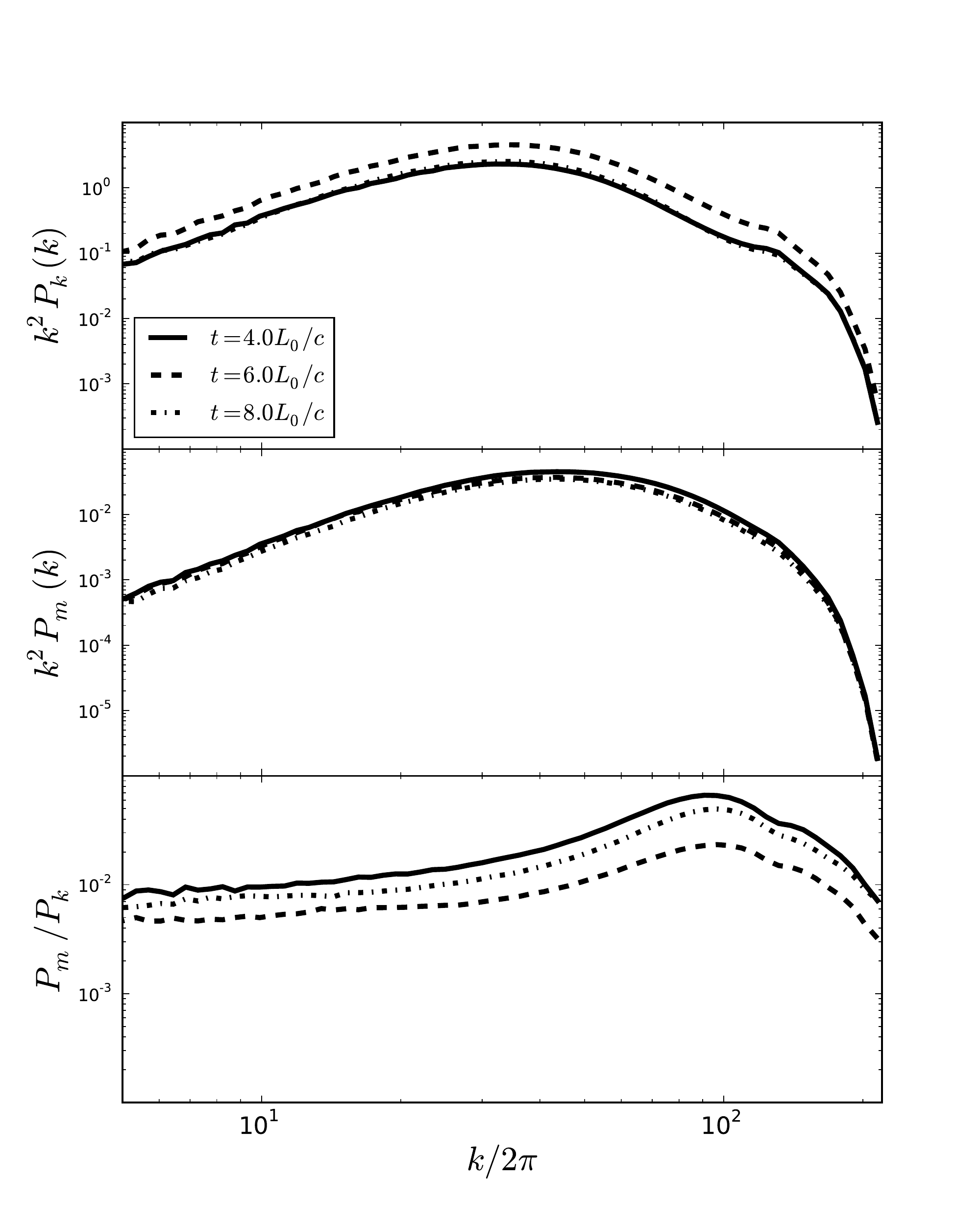}
  \includegraphics[width=3.0in]{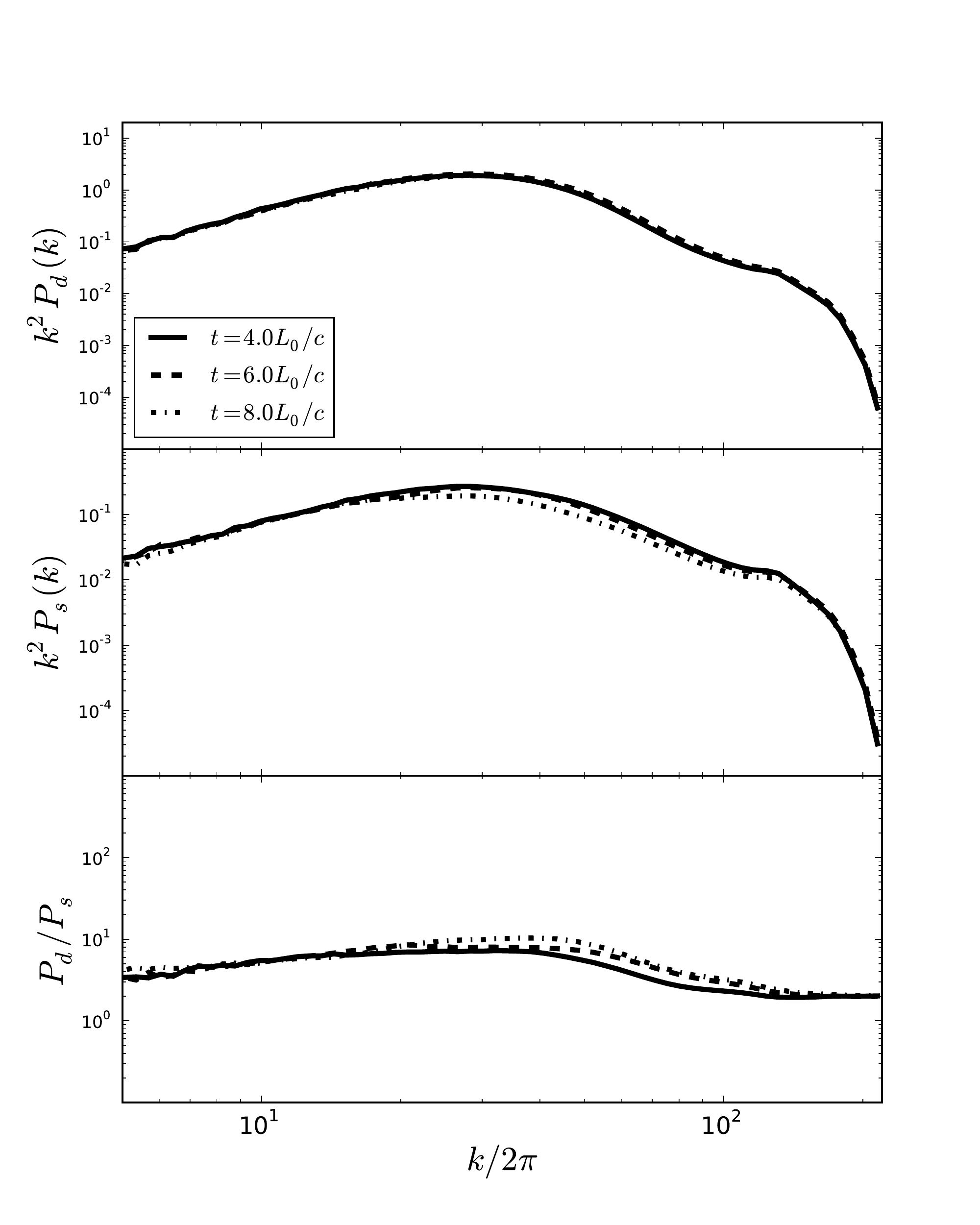}
  \caption{Shown are spherically integrated compensated power spectra of the
    kinetic and magnetic energy (\textbf{left}) and the solenoidal and
    dilatational velocity field modes (\textbf{right}) in a supersonic model
    having $\mathcal{M} \sim 1.7$ at 4 (\emph{solid}), 6 (\emph{dashed}), and 8
    (\emph{dashed-dotted}) light-crossing times.}
  \label{fig:1111-600-512-pspec}
\end{figure}

\section{Time Histories and Power Spectra}
Figure \ref{fig:1111-512-EnergyGrowth} shows the fraction of energy in kinetic,
internal, and magnetic parts from the moment driving is switched on through
saturation and the decay phase. Not shown are the RMS velocities and Mach
number. Within one light-crossing time after driving is switched on the flow
velocities overshoot their steady state values. By three light-crossing times,
they have fallen and the flow has reached a statistically steady state. Note
that the early time transient behavior lasts only until turbulent kinetic energy
has reached the grid scale. The transient may be explained by the absence of the
effective viscosity imposed by smaller scale flow structures. As the energy
cascade reaches smaller scales, this effective viscosity damps the motion of
larger scale eddies, bringing the volume-averaged RMS velocities down. However,
the total kinetic energy does not express the same transient feature and instead
grows monotonically. As kinetic energy populates higher wave-numbers, the
velocity dispersion becomes broader, much like an ensemble of gas particles
being thermalized. However, the velocity structures present at these small
scales are moving less rapidly than the larger eddies in which they are
embedded.

We find that the maximum Lorentz factor of the flow decays exponentially on a
time scale of order the domain's light-crossing time, which is also large eddy
turnover time. The observation that an eddy dissipates its energy to smaller
scales in roughly its own turnover time is in agreement with the conventional
understanding. In particular, it confirms that transitory energy injection at a
given length scale cannot sustain long-lived turbulent eddies at that
scale. This may also be observed in Figure \ref{fig:1111-512-EnergyGrowth} which
shows the rapid decay of kinetic energy after driving is switched off. On the
other hand the volume averaged magnetic energy fraction is sustained in the
absence of driving.

Figure \ref{fig:1111-600-512-pspec} shows spherically integrated power spectra
of the kinetic and magnetic energy, and of the energy in solenoidal and
dilatational velocity field modes. We have computed compenstated power spectra
similar to \cite{Lemaster:2008p2709}, dividing by a power law for ease of
interpretation. We have chosen for simplicity the power law $k^{-2}$, so that
power spectra behaving as a power law with index $-2$ will appear horizontally.


\begin{theacknowledgments}
  This research was supported in part by the NSF through grant AST-1009863 and
  by NASA through grant 09-ATP09-0190.
\end{theacknowledgments}



\bibliographystyle{aipproc}   

\bibliography{ms}

\IfFileExists{\jobname.bbl}{}
{\typeout{}
  \typeout{******************************************}
  \typeout{** Please run "bibtex \jobname" to optain}
  \typeout{** the bibliography and then re-run LaTeX}
  \typeout{** twice to fix the references!}
  \typeout{******************************************}
  \typeout{}
}

\end{document}